# Super-resolution imaging with an achromatic multi-level diffractive microlens array

**Sourangsu Banerji,** [1, ¥] **Monjurul Meem,** [1, ¥] **Apratim Majumder,** [1] **Berardi Sensale-Rodriguez,** [1] **and Rajesh Menon**[1, 2, *]

[1]*Department of Electrical and Computer Engineering, University of Utah, Salt Lake City, UT 84112, USA*
[2]*Oblate Optics, Inc., San Diego, CA 92130, USA*
*\*Corresponding author: rmenon@eng.utah.edu*
*¥denotes equal contribution*

**Compound eyes found in insects provide intriguing sources of biological inspiration for miniaturized imaging systems. Inspired by such insect eye structures, we demonstrate an ultrathin arrayed camera enabled by a flat multilevel diffractive microlens array for super-resolution visible imaging. We experimentally demonstrated that the microlens array can achieve large fill factor (hexagonal close packing with pitch=120μm), thickness of 2.6μm, and diffraction-limited (strehl ratio = 0.88) achromatic performance in the visible band (450nm to 650nm). We also demonstrate super-resolution imaging with resolution improvement of 1.4 times by computationally merging 1600 images in the array.**

Unlike conventional camera lenses, microlenses comparable to insect eye lenses have relatively small focal length as well as low aberrations [1-3], which can substantially reduce the total track length, *i.e.*, the distance from an image sensor to the top surface of the outside lens of a camera [4-6]. These lenses also exhibit larger depth of field (DOF) due to their reduced numerical aperture, which allows near-to-infinity imaging [7]. However, conventional microlenses which consist of stacked refractive lens arrays, still have some technical limitations in decreasing the total track length (TTL) of the camera as they are thicker and suffer from chromatic aberrations [8, 9]. Furthermore, these solutions require precision alignment and integration of multiple optical elements [10]. It is possible to reduce the total track length by utilizing a single achromatic planar and lightweight multilevel diffractive lens (MDL) array and realize the same performance.

Multi-level diffractive lenses (MDLs) are extensions of conventional diffractive lenses that can achieve large operating bandwidths at high efficiencies. Their thickness is typically less ~5*$\lambda$. Since these rely on diffraction, their constituent features are larger and therefore, easier to manufacture. Finally, these lenses exhibit polarization insensitivity, which could be important for imaging. We recently showed that MDLs have equal or better performance when compared to metalenses, which are harder to design and manufacture [11]. MDLs have already been demonstrated in the ultraviolet [12], visible [13], near infrared [14, 15], short wave infrared, [16] long-wave infrared [17], THz [18], and microwave [19] bands. Magnification in an imaging system was also demonstrated by coupling two MDLs [20]. Intriguingly, MDLs have also enabled unusual lens properties such as extremely large bandwidths from 450nm to 15μm [21, 22], large depth of focus [26], and relatively large field of view of ~50° with a single surface [15]. MDLs exploit the basic principle that the phase in the image plane is a free parameter, and thereby, can be adjusted to select the lens pupil transmission function for performance [13].

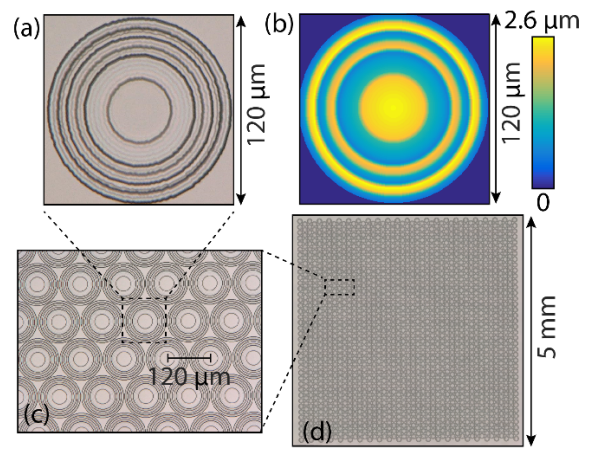

Fig. 1. (a) Optical micrograph of a single fabricated microlens and (b) its corresponding designed geometry. (c) Magnified image of a portion of the hexagonal-closed-packed microlens array with pitch = 120 μm. (d) Image of the entire 40 x 40-microlens array. The size of the array is 5 mm, which is almost of the same size of the CMOS image sensor used in this study.

In this letter, we demonstrate an MDL-based microlens array composed of 40 x 40 microlenses designed for the visible spectrum ($\lambda$ = 450 nm to 650 nm). Each microlens was comprised of 40 concentric rings of width=3µm and heights varying between 0 and 2.6µm (up to 100 gray levels). So, the diameter of the microlens was 120µm and the corresponding focal length was 1mm. The heights of the rings were selected to maximize focusing efficiency (averaged over the bandwidth of interest) as we have described earlier [11]. The focusing efficiency was defined as power within the focused spot [diameter of three times the full width at half-maximum (FWHM)] divided by total incident power. By computing the gradient of this function with respect to the variable ring heights and applying a gradient-directed search, we are able to find the design that is able to maximize the average focusing efficiency.

An optical micrograph of one MDL from the array and the designed geometry are shown in Figs. 1(a) and 1(b), respectively. The device was fabricated in a photoresist (S1813, Microchem) using grayscale optical lithography [13, 21, 22] on top of a double-side polished BK7 glass wafer. Fig 1(c) shows the magnified view of the hexagonal-close-packed array with center-to-center distance (pitch) of 120 µm. The total array size was 5mm (40 x 40 microlenses) as shown in Fig. 1(d). This is almost the same size as the color CMOS image sensor (DFM 72BUC02-ML, Imaging Source) used in this study.

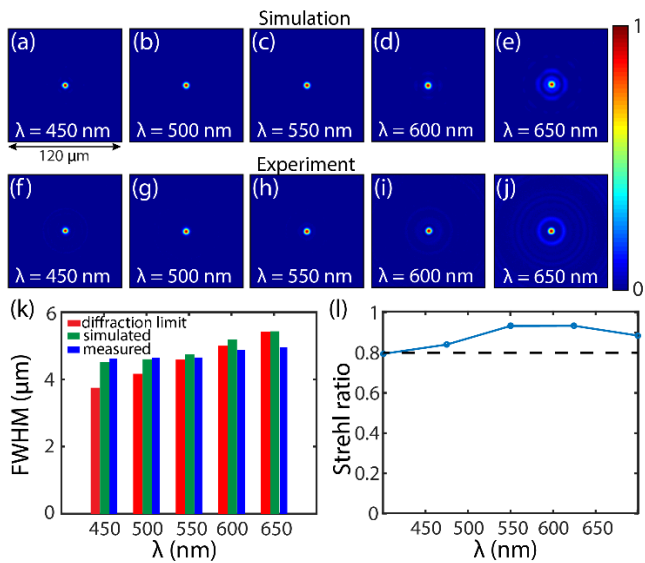

Fig. 2. (a-e) Simulated and (f-g) measured point-spread functions (PSFs) within the designed band: 450-650 nm. (k) Measured, simulated and diffraction-limited full-width at half maximum (FWHM) as a function of wavelength of an individual MDL. (l) Measured strehl ratio (SR) (wavelength averaged SR = 0.88). A lens is considered diffraction limited when the SR > 0.8.

The simulated and measured point-spread functions (PSFs) of one such MDL within the microlens array are shown in Figs. 2(a-e) and 2(f-j), respectively. Each PSF was measured by illuminating the MDL with collimated light from a supercontinuum source coupled to tunable filters [13]. The focal plane of the MDL was then magnified using an objective (RMS20X-PF, Thorlabs) and tube lens (ITL200, Thorlabs) and imaged onto a monochrome CMOS image sensor (DMM 27UP031-ML, Imaging Source). The gap between the objective and tube lens was ~90 mm and that between the sensor and the backside of the tube lens was ~148 mm. The magnification of the objective–tube lens was 22.22x. This was necessary to capture the diffraction limited focal spot size at the focal plane with sufficient spatial resolution. The theoretical diffraction-limited, simulated and the measured FWHM as function of $\lambda$ are shown in Figs. Fig. 2(k). The wavelength-averaged theoretical diffraction limited, simulated and measured FWHM values are 4.58 µm, 4.78 µm and 4.75 µm respectively. The plot of the Strehl Ratio (SR) was measured and plotted as function of $\lambda$ in Fig. 2(l). Since the SR > 0.8 over the entire visible band (average value is 0.88), the MDL is clearly diffraction limited and achromatic.

We calculated the wavefront errors as linear sums of Zernike polynomials based on the simulated PSF at $\lambda$ = 550 nm (Fig. 3a). The aberrations were defined as the difference between the measured wavefront and the ideal spherical wavefront, and the difference is expressed as the coefficients of the Zernike polynomials. Fig. 3(a) shows the coefficients for the first 15 types of geometric aberrations, which confirms very low aberrations.

Next, we calculated the modulation-transfer function (MTF) of the MDL by taking the absolute value of the Fourier transform of each recorded PSF from Figs. 2(f-j). The MTF summarized in Fig. 3(b) confirms an average value of 175 lp/mm at 10% contrast.

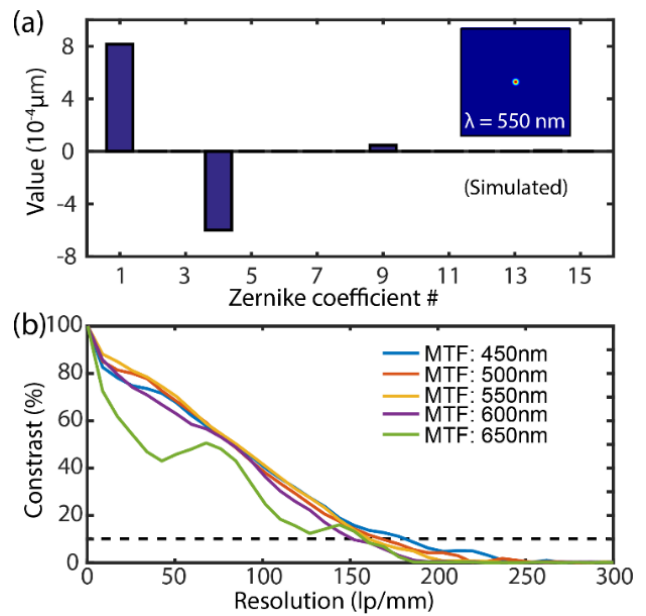

Fig. 3. (a) Aberrations analysis from simulated PSF at $\lambda$=550nm (inset). (b) Experimental modulation-transfer-function (MTF). Resolution is 175 lp/mm at 10% contrast.

Finally, we assembled an ultrathin arrayed camera (Fig. 4a) by placing the microlens array ~1mm away in front of the image sensor. Next, an object was placed at a distance of 350mm from the camera and illuminated with a white LED flashlight. The recorded image is shown in Fig. 4b. Each image in this array was extracted and registered to a single coordinate system as described by Keren et. al. [24]. Finally, these images were merged to reconstruct a super-resolved image. The merging procedure followed he "anisotropic diffusion super resolution" method proposed by Kim et. al. [25]. Figure 4(c) shows the reconstructed images as a function of the number of constituent raw images. The reference image taken with a Google Pixel 3 camera is shown in Fig. 4(d) and the best reconstructed image (1600 merged images) is shown in Fig. 4(e). To demonstrate super-resolution quantitatively, we computed MTF via a slanted-edge method processed through open source software (MTFMapper) of a single microlens image and the final reconstructed image [26]. The resolution at 10% MTF contrast was improved by a factor of 1.4 X (to ~245 lp/mm). In fig

4(g), we imaged a scene comprised of 3 jellybeans at an object distance of 250mm (image distance ~ 1mm) to emphasize the perspective shift from one edge of the microlens array to the other.

In summary, we demonstrated a hexagonal-close-packed-microlens array comprised of MDLs with active thickness of 2.6μm, pitch of 120μm and focal length of 1mm. The measured PSFs, MTFs and strehl ratio confirm achromatic diffraction-limited performance across the visible spectrum. By computationally merging all 1600 images in the array, we also demonstrated super-resolution by increasing the resolved line-pairs at 10% contrast from 175 lp/mm to 245 lp/mm over the entire visible band.

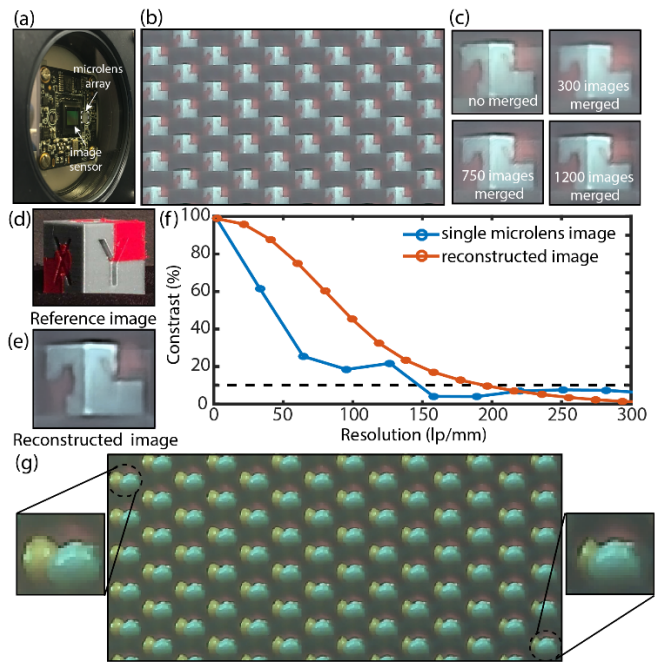

Fig. 4. Imaging with the microlens array. (a) Photograph of assembled camera. (b) Raw image of a cube with object distance = 330mm and image distance ~ 1mm. (c) Reconstructed super-resolved images as function of number of raw images used. (d) Reference image captured by the Google Pixel 3 camera. (e) Final reconstructed super-resolved image using 1600 microlens images. (f) Improvement in MTF demonstrating super-resolution by about 1.4X. (g) Raw image of 3 jellybeans with object distance = 250mm and image distance ~ 1mm to observe the perspective shift across the array.

**Funding.** Office of Naval Research grant N66001-10-1-4065 NSF: 1351389, 1828480, and 1936729.

**Acknowledgments**. We thank Brian Baker, Steve Pritchett, and Christian Bach for fabrication advice, and Tom Tiwald (Woollam) for measuring the dispersion of materials. The support and resources from the Center for High Performance Computing at the University of Utah are also gratefully acknowledged.

**Disclosures.** Rajesh Menon is the founder of Oblate Optics, Inc., which is commercializing technology discussed in this Letter. The University of Utah has filed for patent protection for technology discussed in this Letter.